\documentclass[useAMS,usenatbib,usegraphicx]{mn2e}

\newcommand{\kev}{keV}
\newcommand{\fe}{Fe~K$\alpha$}
\newcommand{\etal}{et al.}

\title[Fe~K$\alpha$ from ionized slabs: the impact of the iron abundance]
  {Fe~K$\mathbf{\alpha}$ emission from photoionized slabs: the impact
  of the iron abundance}
\author[D.\ R.\ Ballantyne, A.\ C.\ Fabian \& R.\ R.\ Ross]
  {D.~R.~Ballantyne$^1$\thanks{drb@ast.cam.ac.uk}, A.~C.~Fabian$^1$
  and R.~R.~Ross$^2$\\
  $^1$Institute of Astronomy, Madingley Road, Cambridge CB3 0HA \\
  $^2$Physics Department, College of the Holy Cross, Worcester, MA 01610, USA}

\pagerange{\pageref{firstpage}--\pageref{lastpage}}
\pubyear{2001}

\usepackage{times}

\begin{document}

\label{firstpage}

\maketitle

\begin{abstract}
Iron K$\alpha$ emission from photoionized and optically thick material
is observed in a variety of astrophysical environments including X-ray
binaries, active galactic nuclei, and possibly gamma-ray bursts. This
paper presents calculations showing how the equivalent width (EW) of
the \fe\ line depends on the iron abundance of the illuminated gas and
its ionization state -- two variables subject to significant cosmic
scatter. Reflection spectra from a constant density slab which is
illuminated with a power-law spectrum with photon-index $\Gamma$ are
computed using the code of Ross \& Fabian. When the \fe\ EW is
measured from the reflection spectra alone, we find that it can reach
values greater than 6~\kev\ if the Fe abundance is about 10$\times$
solar and the illuminated gas is neutral. EWs of about 1~\kev\ are
obtained when the gas is ionized. In contrast, when the EW is measured
from the incident+reflected spectrum, the largest EWs are $\sim
800$~\kev\ and are found when the gas is ionized. When $\Gamma$ is
increased, the \fe\ line generally weakens, but significant emission
can persist to larger ionization parameters. The iron abundance has
its greatest impact on the EW when it is less than 5$\times$
solar. When the abundance is further increased, the line strengthens
only marginally. Therefore, we conclude that \fe\ lines with EWs much
greater than 800~eV are unlikely to be produced by gas with a
supersolar Fe abundance. These results should be useful in
interpreting \fe\ emission whenever it arises from optically thick
fluorescence.
\end{abstract}


\section{Introduction}
\label{sect:intro}

There are a number of different astrophysical environments where
optically thick material may be irradiated by X-rays. If the incident
spectrum has significant flux above 7.1~\kev, then, as a result of its
relatively high cosmic abundance and large fluorescent yield, the
iron~K$\alpha$ line is predicted to be a significant feature in the
resulting emission from the illuminated surface
\citep*[e.g.,][]{gf91,mpp91}. The presence of this line has been
discussed for a number of different reprocessors including   
 the solar surface \citep{bai79}, the surface of the
companion star in X-ray binaries \citep*{bst74,b78,p79}, and the
surface of accreting magnetic white dwarfs
\citep*[e.g.,][]{sfr84,dob95,vkh96}. Most famously, \fe\ emission has
been observed in the X-ray spectra of many Active Galactic Nuclei
(AGN) and black hole candidates \citep*{np94,eb96}. Here, the
irradiated surface is likely the accretion disc feeding the black
hole, and some of the observed \fe\ lines have been observed to be
broadened in a manner consistent with material orbiting in a
relativistic gravitational potential \citep*{tan95,nan97,fab00}. Finally,
tentative detections of line emission, most likely from \fe, have been
found in recent observations of the X-ray afterglows of some
$\gamma$-ray bursts \citep*[e.g.,][]{pir99,pir00,ant00}. If this is
confirmed then it implies that optically thick material may be in the
vicinity --- a possible constraint on the environment and progenitor
of the burst \citep*{ree00,vie01,brr01}.

As one of the few physical diagnostics in the X-ray spectra of these
systems, it is important to understand how the \fe\ emission line
depends on the properties of the irradiated gas. Previous studies have
examined the influence of changes in metal abundance \citep*{mfr97},
the inclination angle of the reflector \citep*{ghm94}, ionization
state \citep*{mfr93,mfr96}, and the density distribution of the gas
\citep*{ros99,nay00,br01} on the line emission. This paper generalizes
the results of Matt and colleagues by presenting how the strength of the
\fe\ line (presented as an equivalent width [EW]) depends on both the
Fe abundance and the ionization state of the gas. These two variables
have the greatest impact on the EW, and are likely to vary widely over
different astrophysical environments. We anticipate that these results
will be helpful in interpreting \fe\ detections whenever it arises
from an optically thick surface.

The following section details the calculations that were performed,
and then Section~\ref{sect:res} presents the results as contour plots
of EW. A brief discussion concludes the paper in
Section~\ref{sect:discuss}.

\section{Computations}
\label{sect:comp}

The calculations were performed using the code of \citet{ros93}. A
slab with a constant hydrogen number density
$n_{\mathrm{H}}=$10$^{15}$~cm$^{-3}$ was illuminated with a power-law
continuum with photon-index $\Gamma$ (i.e., so that photon flux
$\propto E^{-\Gamma}$). Provided that they have comparable flux above
7~\kev, other incident spectral forms (e.g., blackbody, Wien tail)
will not significantly alter the results. As a crude approximation to
isotropic illumination, the penetrating radiation was assumed to
strike the slab at an angle of 54.7~degrees from the normal. Hydrogen
and helium were assumed to be fully ionized, but C~{\sc v--vii},
O~{\sc v--ix}, Mg~{\sc ix--xiii}, Si~{\sc xi--xv}, and Fe~{\sc
xvi--xxvii} were treated with abundances given by \citet{mcm83}. The
depth of the slab was increased as the illuminating flux was
increased, and had a minimum value of 6 Thomson depths and a maximum
of 30 Thomson depths.

When the irradiated material reached thermal and ionization
equilibrium, the angle-averaged reprocessed emission from the gas was
calculated. As with all photoionized gas, the reflection spectrum can
be characterized by an ionization parameter:
\begin{equation}
\label{eq:ip}
\xi = {4 \pi F_{\mathrm{X}} \over n_{\mathrm{H}}},
\end{equation}
where $F_{\mathrm{X}}$ is the incident flux between 0.01 and 100~\kev,
the energy range of the illuminating continuum. $\xi$ was varied by
changing only the value of $F_{\mathrm{X}}$. The evolution of the
reflection spectrum with $\xi$ has been discussed by
\citet{ros99}. Here we calculate reflection spectra for $\log
\xi=1.0$--6.0 while also varying the Fe abundance from 0.1--10.1 times
solar. When the illuminated gas is predominantly neutral, the \fe\
line may be affected by variations in other metals, particularly
oxygen \citep{mfr97}. Of course, when $\log \xi \ga 2.8$ all the
metals other than Fe are fully ionized and will have little effect on
the Fe emission lines. Therefore, we do not consider abundance effects
due to other metals, and instead refer readers to \citet{mfr97} where
such results are presented for neutral reflectors. The iron emission
features will also depend slightly on the slope of the incident
continuum (due to the different number of ionizing photons), so we
have performed these computations for $\Gamma=1.7$, 2.0 and 2.3 to
illustrate this effect.

\begin{figure}
\includegraphics[width=0.50\textwidth]{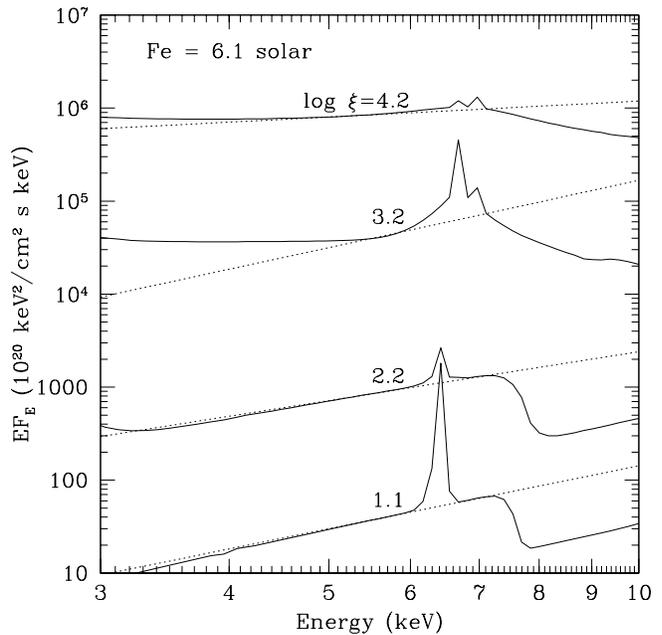}
\caption{Illustration of how the equivalent widths were computed
from the calculated reflection spectra where, in this case, a
$\Gamma=2.0$ continuum was incident on the slab. The solid lines show
the reprocessed emission and the dotted lines are the estimate to the
local continuum about the \fe\ line. The EW of the line is then
calculated by integrating the flux above the dotted line between 5.5
and 7.1~\kev.}
\label{fig:ew-ill-noR}
\end{figure}

The EWs were calculated by the following integral:
\begin{equation}
\label{eq:ew}
\mathrm{EW} = \int_{5.50~\mathrm{keV}}^{7.11~\mathrm{keV}} { EF_s(E) - EF_c(E)
\over EF_c(E)} dE,
\end{equation}
where $EF_s(E)$ is the spectral flux of the reflected spectrum, and
$EF_c(E)$ is the estimated spectral flux of the continuum at energy
$E$. The continuum was estimated by fitting a straight line to the
reflection spectrum between 5.50 and 7.11~\kev. The lower limit of the
integration was chosen to take into account any Compton broadening of
the line profile, and the upper limit was set to the energy of the
photoelectric absorption edge of neutral Fe. Dividing the line-flux by
the monochromatic continuum at 6.4 or 6.7~\kev\ makes a negligible
difference to the contour plots. Figure~\ref{fig:ew-ill-noR} shows
several examples of this calculation at different $\xi$.

\section{Results}
\label{sect:res}

\begin{figure*}
\begin{minipage}{180mm}
\centerline{
\includegraphics[width=0.57\textwidth]{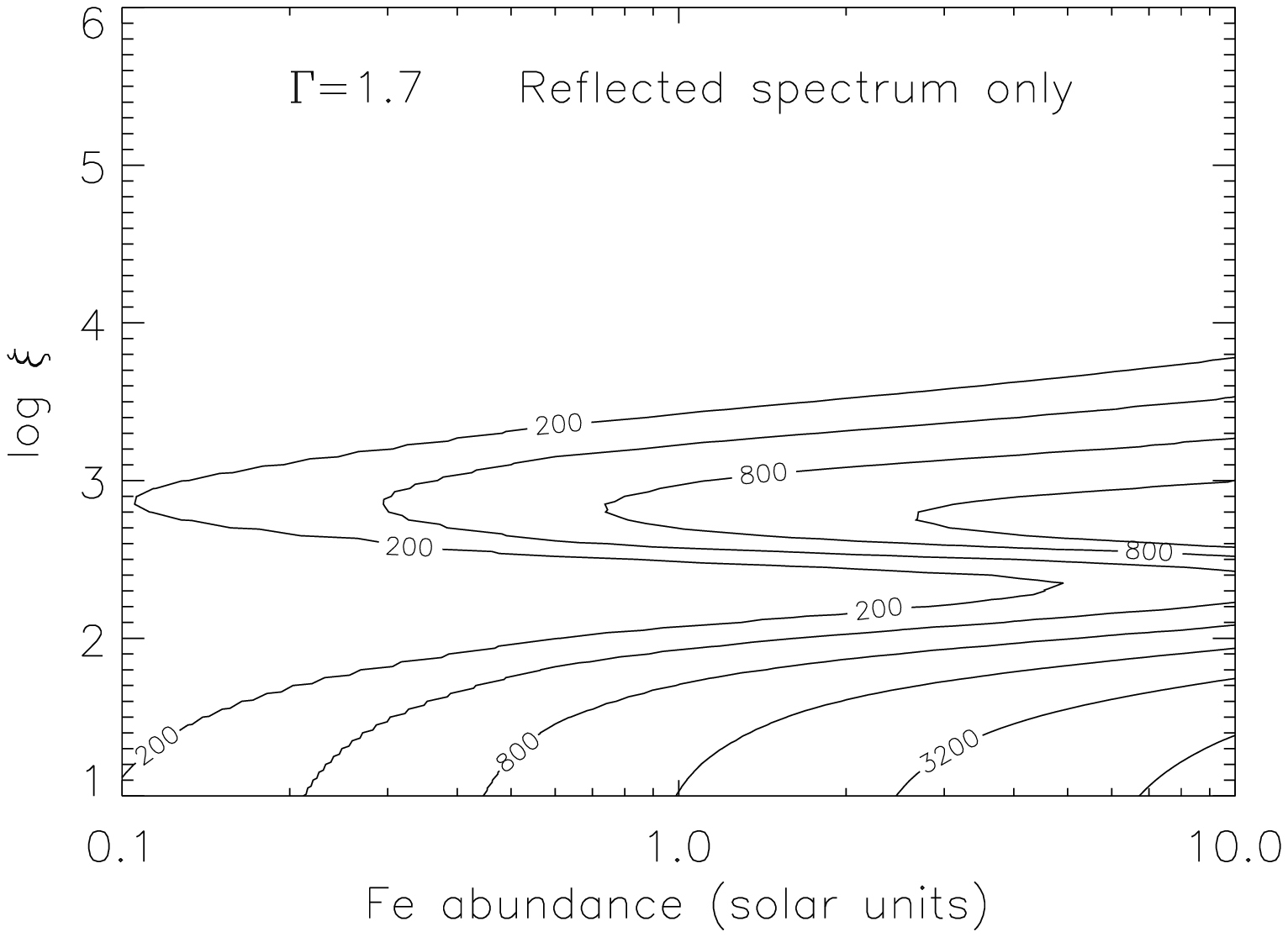}
\includegraphics[width=0.57\textwidth]{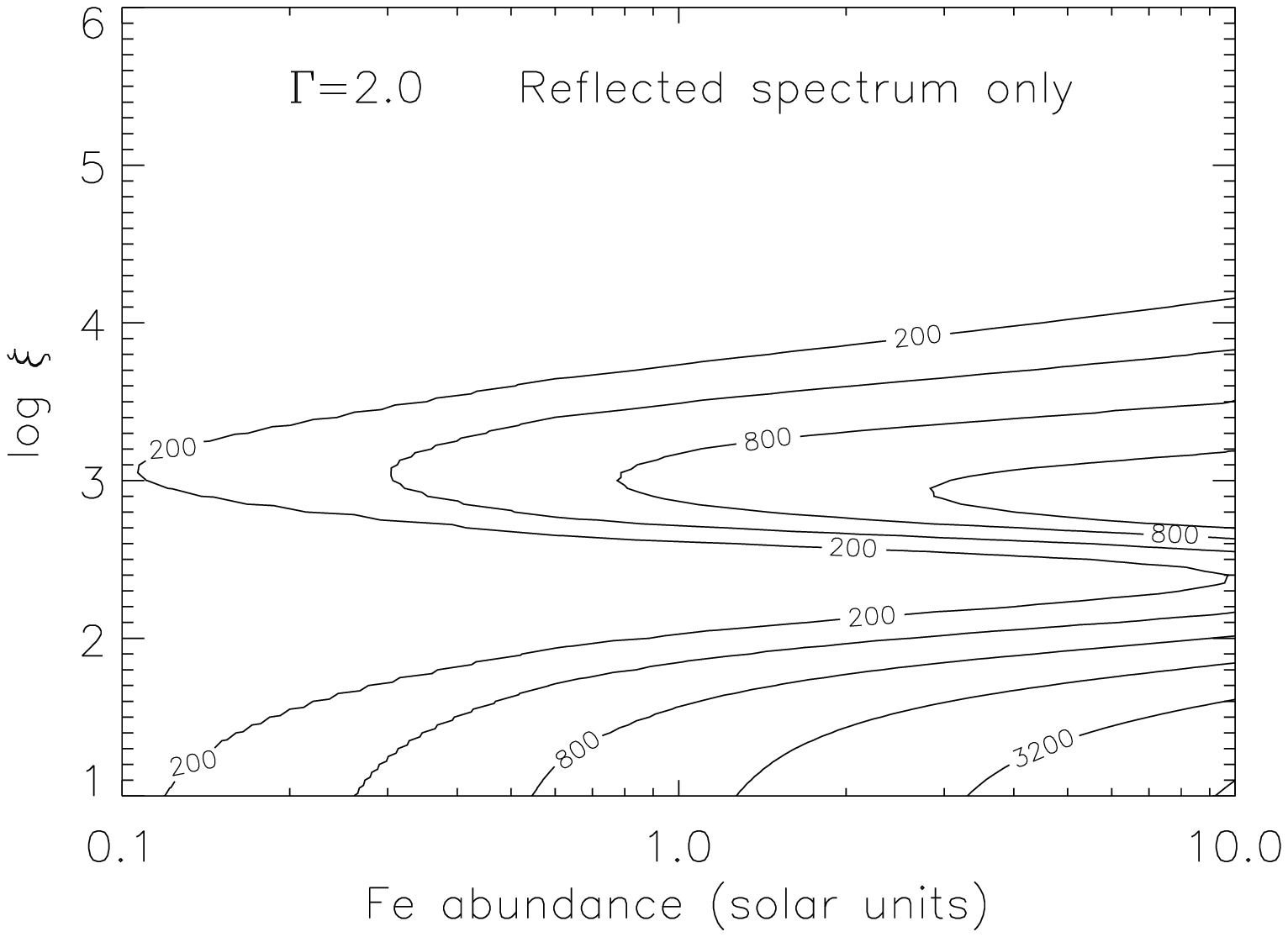}
}
\vspace{0.1mm}
\centerline{
\includegraphics[width=0.57\textwidth]{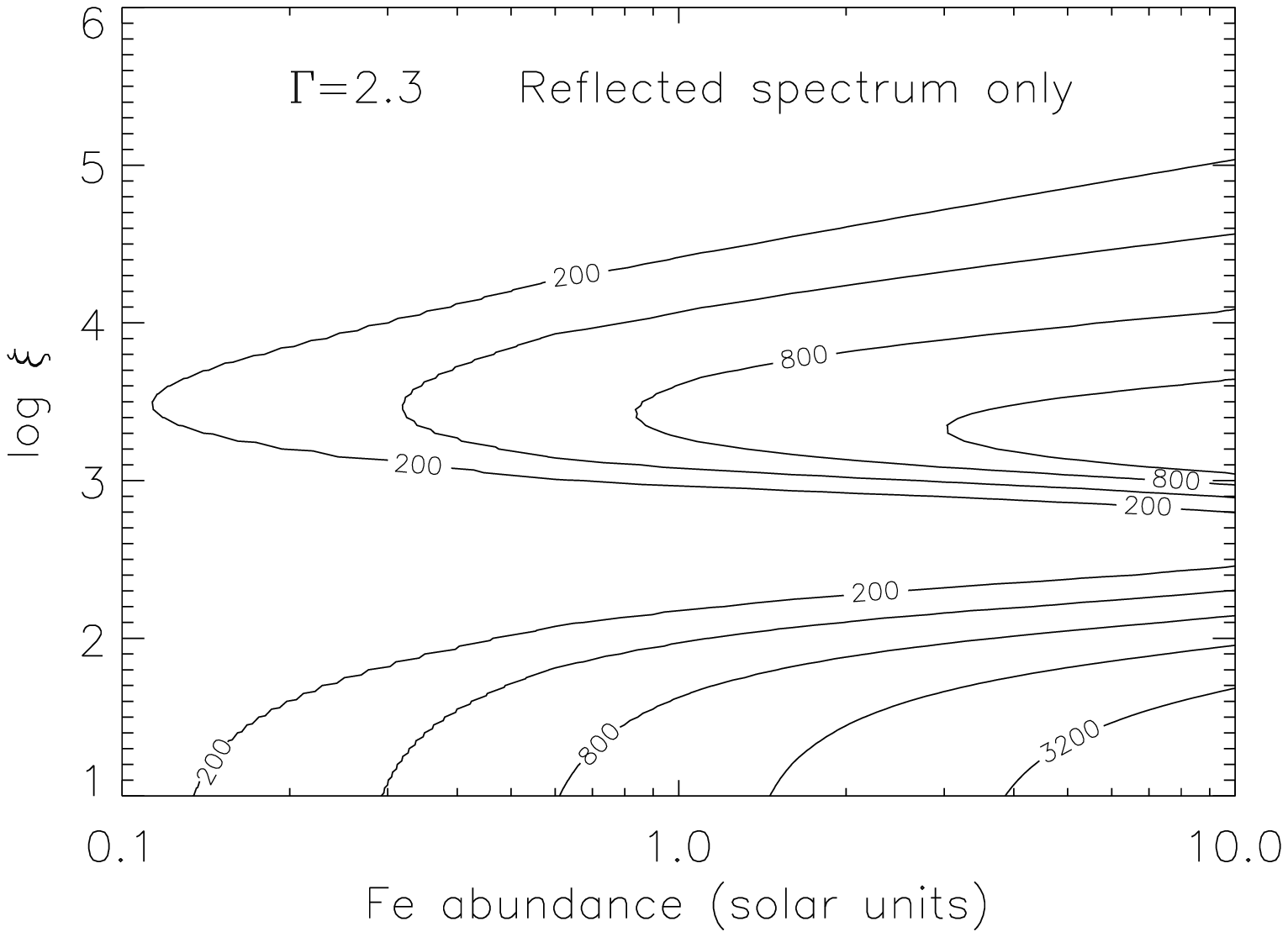}
}
\caption{Each panel shows contours of constant \fe\ EW in the $\log
\xi$-Fe abundance plane. The different panels separate the three
different values of $\Gamma$ that were considered. The contours are at
200, 400, 800, 1600, 3200, 6400~eV and notice that only every second
contour is labeled. The incident power-law was not added to the
reflection spectra before the line was measured, therefore these
EWs are the maximum possible under these circumstances.}
\label{fig:contourplots-noR}
\end{minipage}
\end{figure*}

Figure~\ref{fig:contourplots-noR} shows contours of constant \fe\ EW
in the $\log \xi$-Fe abundance plane for the case where the incident
power-law has not been added to the reflected spectra, as if the
reflected spectrum were viewed in isolation. Therefore, these EWs are
the \textit{maximum} that can be observed as they are not diluted by
the incident power-law. Such reflection-dominated spectra may be
observed in Seyfert~2 galaxies (e.g., NGC~6552; \citealt{rfm94}).
The plots show that the \fe\ EW reaches two separate maxima in the
vertical direction: one when the ionization parameter is low and the
line is at 6.4~\kev, and the other at $\log \xi \sim 3$ when the line
is at 6.7~\kev. Between these two maxima the line is suppressed due to
Auger destruction \citep*{rfb96}, and at higher $\xi$ the line weakens and
disappears as all the iron becomes ionized. 

Increasing the iron abundance does increase the \fe\ EW, but, as is
discussed in Sect.~\ref{sect:discuss}, its effect is greatest when
the abundance is still relatively low. Nevertheless,
Fig.~\ref{fig:contourplots-noR} shows that the line can get very
strong ($> 6$~\kev\ in some cases) at high Fe abundance with the
largest values occurring when the line is neutral. As is shown
below, this is reversed when the incident power-law is added to the
reflection spectrum.

As $\Gamma$ is increased the number of Fe ionizing photons
decreases. This causes the contours to spread out vertically in the
plot, because at a higher $\Gamma$ a larger flux is needed to ionize
Fe. When $\log \xi \la 2.5$ and $\Gamma > 1.7$ the dependence on the
Fe abundance weakens and the EW grows more slowly. Again, this is due
to the decrease in the number of ionizing photons.

\begin{figure*}
\begin{minipage}{180mm}
\centerline{
\includegraphics[width=0.57\textwidth]{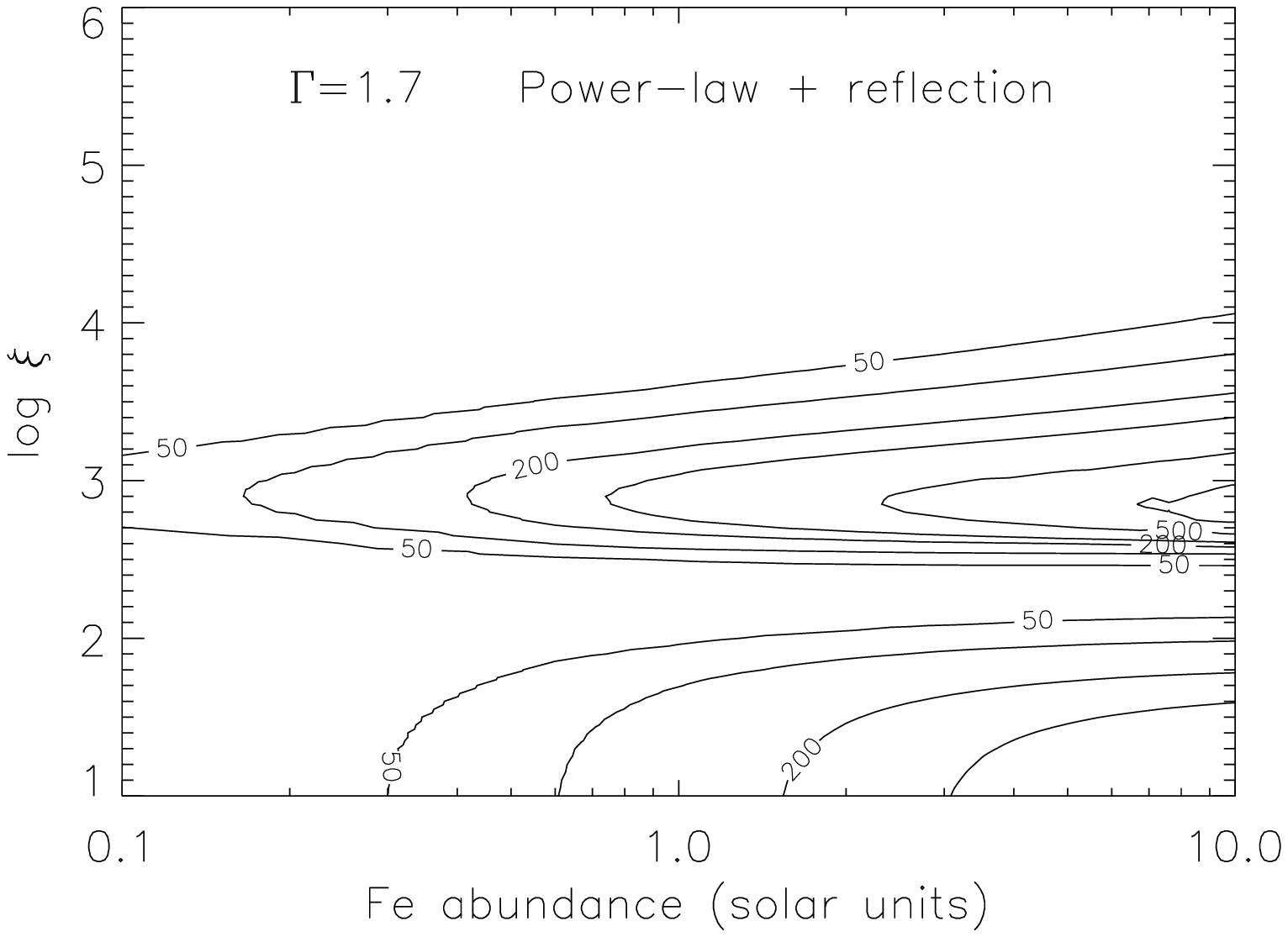}
\includegraphics[width=0.57\textwidth]{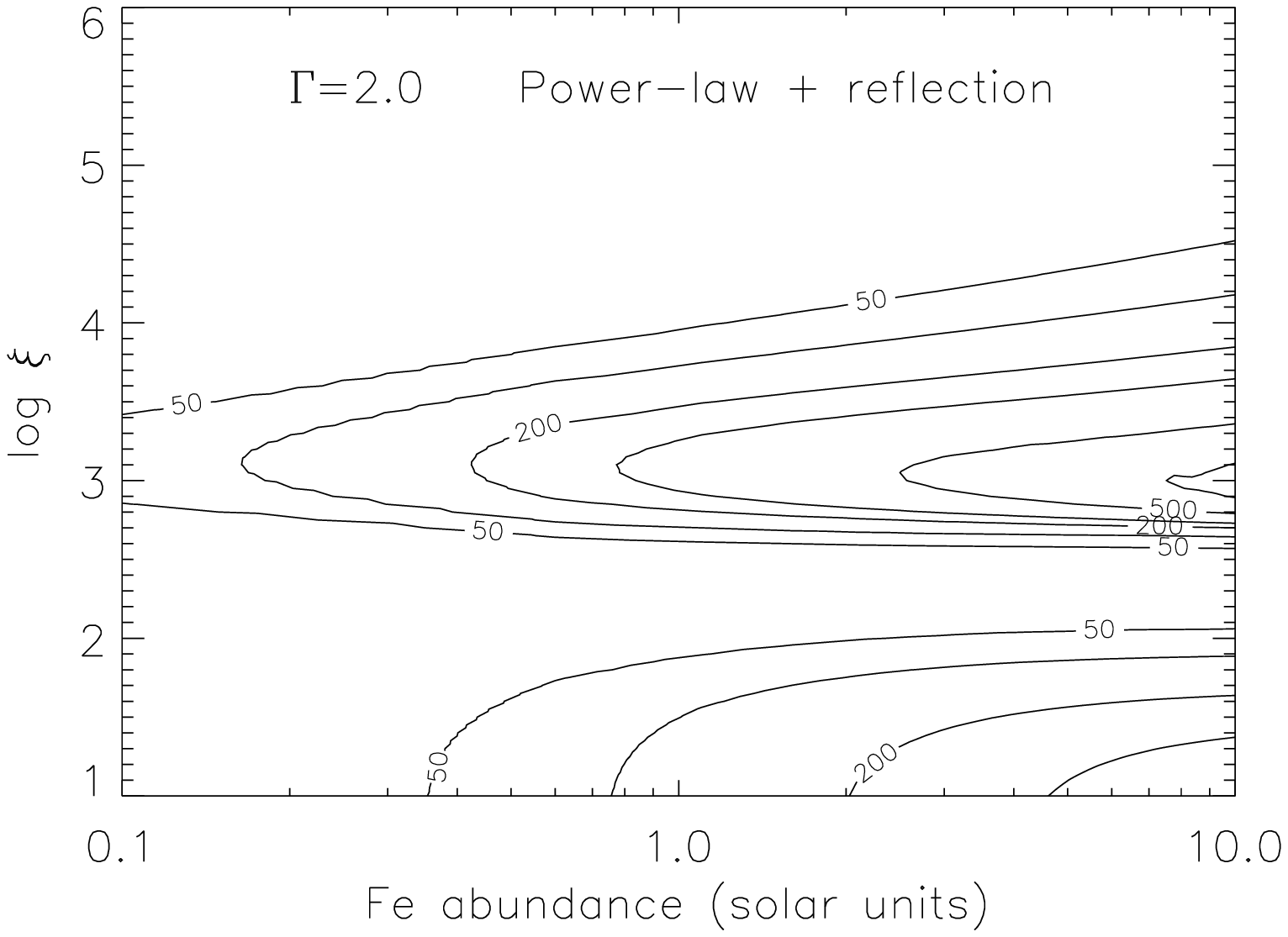}
}
\vspace{0.1mm}
\centerline{
\includegraphics[width=0.57\textwidth]{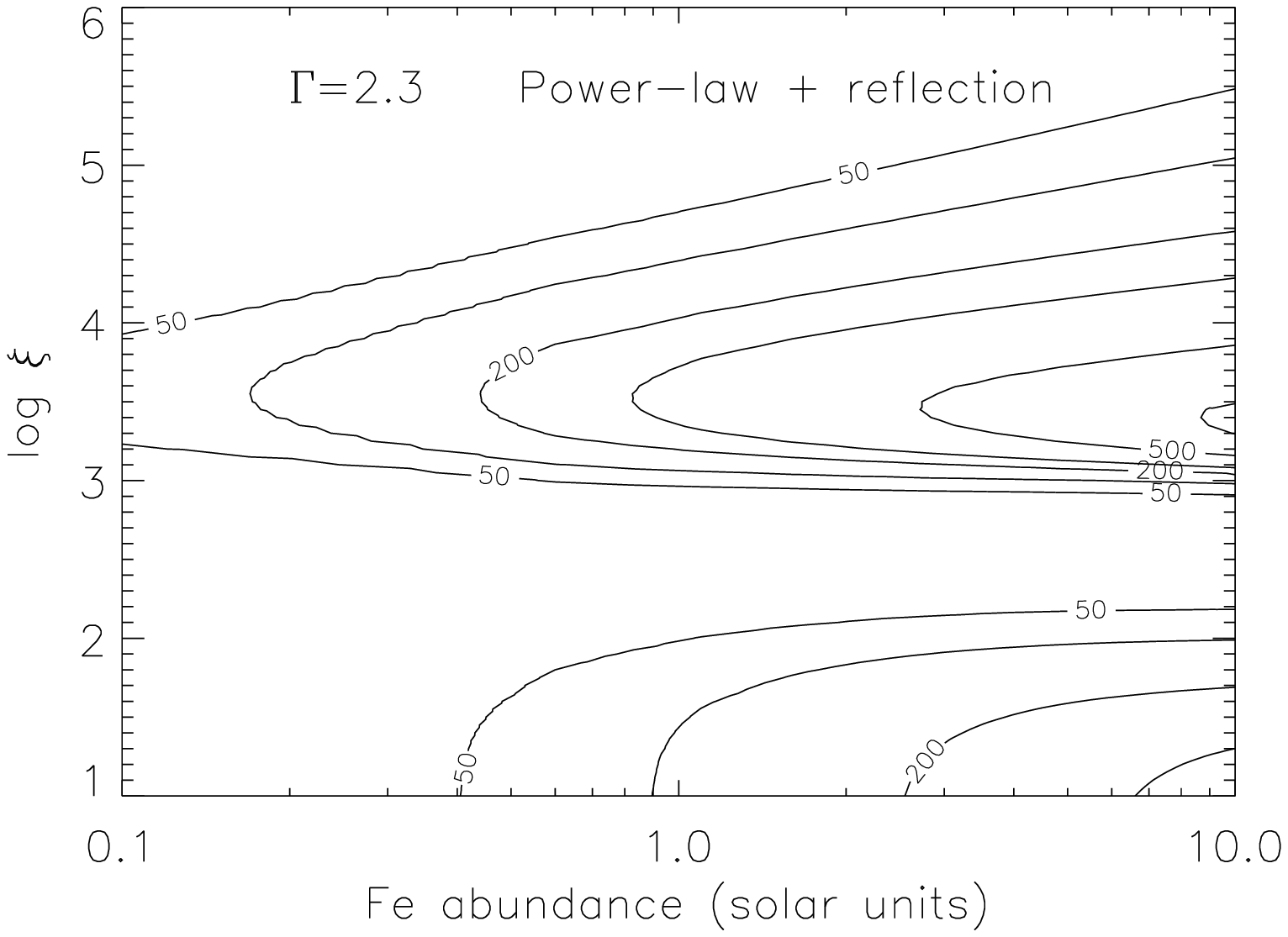}
}
\caption{Each panel shows contours of constant \fe\ EW in the $\log
\xi$-Fe abundance plane. The different panels separate the three
different values of $\Gamma$ that were considered. The contours are at
50, 100, 200, 300, 500, 700~eV and notice that only every second
contour is labeled. In this case, the incident power-law \textit{was}
added to the reflection spectra before the line was measured.}
\label{fig:contourplots}
\end{minipage}
\end{figure*}

In many X-ray observations unprocessed radiation with the spectrum of
the incident power-law is simultaneously detected along with the
reflection spectrum. Therefore, we have repeated all the EW
calculations for the total power-law+reflection spectra (AGN workers
think of this as a reflection fraction of unity). These results are
shown in Figure~\ref{fig:contourplots}.
As before, two maxima are found in the contours: one when iron is
neutral and the other when the He-like \fe\ emission line at 6.7~\kev\
is prominent. Likewise, the contours spread out vertically as the
photon-index is increased. However, there are a number of
differences between this set of plots and the previous one. The first
one is trivial but important: the \fe\ EWs are now much smaller. The
largest values are now just over 800~eV and occur when $\log \xi \sim
3$. At low values of $\xi$ the 6.4~\kev\ line dominates, and the EW
cannot exceed 400~eV. This turn-around between the (apparent) strength
of the neutral and ionized lines when the incident spectrum is added
in is caused by the effects of absorption by other elements in the
gas. When the slab is weakly illuminated much of the flux around
6~\kev\ is absorbed by other metals in the gas, particularly oxygen,
and is re-emitted as softer X-rays. This results in a large difference
in flux between the incidence spectrum and the reflected spectrum at
6~\kev. In contrast, when $\log \xi \sim 3$ all the metals except for
iron are almost completely ionized, so there is very little absorption
around the \fe\ line. Thus, when the reflection and incident spectra
are added together, the He-like 6.7~\kev\ line is more prominent
than the neutral line.

To summarize, when iron abundances up to 10.1$\times$ solar were
considered, the maximum \fe\ EW is about 8.5~\kev\ and is found when
the gas is neutral and illuminated by $\Gamma=1.7$ power-law. In the
case when the power-law is just simply added to the reflection
spectrum (i.e., a reflection fraction of 1), the EWs are much smaller ($<
1$~\kev) with the largest values being found when the gas is ionized
and the 6.7~\kev\ line of He-like Fe is dominant. In this case, it
seems to be very difficult to obtain \fe\ lines with EWs greater than
700~eV.  

\section{Discussion}
\label{sect:discuss}

The results presented in the previous section showed that increasing
the iron abundance would increase the \fe\ EW, but its influence seemed to
weaken as it grew. It is interesting to investigate this point since
we have only considered abundances up to 10.1$\times$ solar. Is it
possible to obtain an arbitrarily large EW solely by increasing the Fe
abundance? 

\begin{figure}
\centerline{
\includegraphics[width=0.56\textwidth]{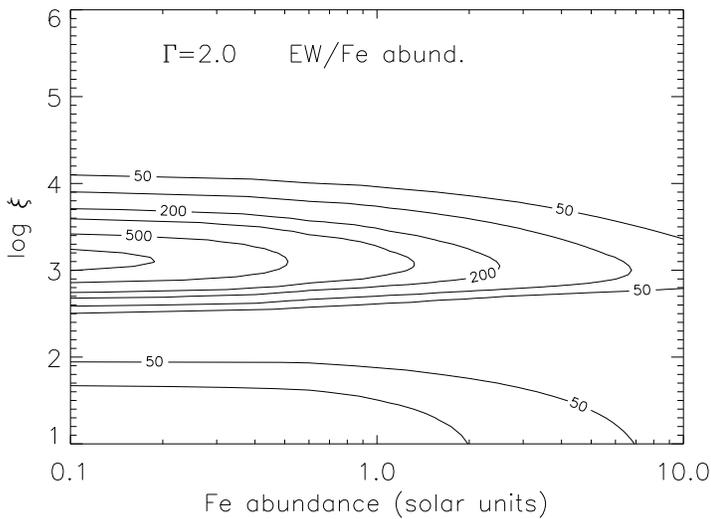}
}
\caption{Contours of (\fe\ EW/Fe abundance) in the $\log
\xi$-Fe abundance plane. The EWs were calculated using the total
incident+reflected spectra and we only show the case of $\Gamma=2.0$.
The contour lines are at 50, 100, 200, 300, 500, 700~eV. The effect of
increasing the iron abundance is greatest when it is less than
3$\times$ solar and when the slab is ionized.}
\label{fig:gamma2-perion}
\end{figure}

The answer to this question is found in Figure~\ref{fig:gamma2-perion}.
This plot shows contours of \fe\ EW normalized by the Fe abundance for
the case of $\Gamma=2.0$ and a reflection fraction of one. In this
case the contours show a maximum at low Fe abundance implying that the
most ``power'' out of an Fe atom or ion occurs when its abundance is
around solar. Increasing the amount of iron in the gas does permit the
number of \fe\ photons emitted to increase, but it also strengthens
the line destruction mechanisms of photoabsorption and
scattering. Therefore, for a given flux of ionizing photons, competing
atomic processes in the gas eventually decreases the efficiency of
iron fluorescence. This is seen particularly when the gas is neutral:
the normalized \fe\ EW falls slowly over a wide range in Fe abundance
because K$\alpha$ photons are destroyed by L shell absorption. 

\begin{figure}
\includegraphics[width=0.50\textwidth]{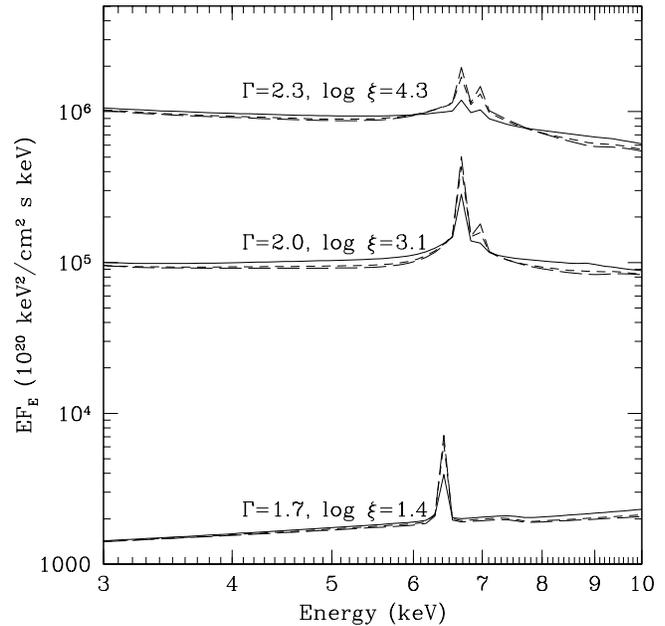}
\caption{Explicit examples of how the \fe\ line changes with Fe
abundance. Total incident+reflection spectra are plotted for the three
different values of $\Gamma$ and three representative values of
$\xi$. Reflection from a slab with a 1.1, 5.1 and 10.1$\times$ solar
Fe abundance is plotted as the solid, short-dashed and long-dashed
lines, respectively.}
\label{fig:line-evolution}
\end{figure}

At higher values of the ionization parameter, there is a more rapid
drop in the normalized \fe\ EW. As is illustrated explicitly in
Figure~\ref{fig:line-evolution}, 
an increase in the iron abundance from 1.1 to 5.1$\times$ solar causes
a fairly big jump in the line strength, but then only a very small
change from 5.1 to 10.1$\times$ solar. In this case, only K shell
absorption is important, but due to the finite number of ionizing
photons, the K edge will saturate when the Fe abundance is about solar
(from Fig.~\ref{fig:gamma2-perion}). Increasing the Fe abundance
beyond this point will not greatly increase the line emission because
there are few photons available to ionize the newly introduced Fe
atoms. Of course, increasing $\xi$ allows more K shell ionizations to
occur, and the normalized EW becomes more uniform with Fe
abundance. Interestingly, the H-like \fe\ line at 6.97~\kev\ which is
suppressed due to resonance scattering significantly strengthens as
the abundance is increased.

These results illustrate that above an Fe abundance of $\sim$
5$\times$ solar there is only a very slow increase in the \fe\
EW. Therefore, it seems unlikely that the EW of the line can be
significantly increased by considering iron abundances larger than
10$\times$ solar.  

\section*{Acknowledgments}
DRB acknowledges financial support from the Commonwealth Scholarship
and Fellowship Plan and the Natural Sciences and Engineering Research
Council of Canada. ACF and RRR acknowledge support from the Royal
Society and the College of the Holy Cross, respectively. ACF and RRR
thank A.\ Young for his work on this project.


\bsp 

\label{lastpage}

\end{document}